\documentclass[aps,pra,onecolumn,groupedaddress,showpacs]{revtex4}

\usepackage{amssymb,amsmath}
\usepackage{graphicx}

\newcommand{\om}{\omega}

\newcommand{\pa}{\partial}
\DeclareMathOperator{\sech}{sech}

\begin{document}

\title{Two-component nonlinear wave solutions of the sixth-order generalised Boussinesq-type equations}

\author{G. T. Adamashvili}
\affiliation {Technical University of Georgia, Kostava str.
77, Tbilisi, 0179, Georgia. }

\begin{abstract}

Two different versions of cubic sixth-order generalised Boussinesq-type wave equations are considered in this study. A generalised perturbation reduction method is used to solve these equations, which allows the reduction of considered equations to coupled nonlinear Schr\"odinger  equations. Two-component nonlinear wave solutions are obtained. The profiles and parameters of these solutions for both nonlinear equations are presented and compared. These solutions coincide with the vector $0\pi$ pulse of self-induced transparency, which was previously studied in several known nonlinear wave equations.
\end{abstract}

\pacs{42.50.Md; 42.65.Tg;  77.22. Gm}

\maketitle

\section{Introduction  }

Among physical phenomena of particular interest are those that occur in completely different physical systems and consequently express the most general properties of these systems. Such phenomena are important because they characterise the most fundamental properties of matter, regardless of their physical nature. Such phenomena belong to the existence of nonlinear solitary waves, which occur in almost all physical systems [1-9]. Among nonlinear solitary waves, single-component and two-component nonlinear waves are mainly distinguished. The first proof of the existence of single-component solitary nonlinear waves is associated with the analysis of the Boussinesq equation. This equation describes the propagation of small amplitude long waves on the surface of shallow water [10]. Many works have investigated the single-component solitary nonlinear wave solution of the Boussinesq equation analytically and numerically using different mathematical methods for a very long time. Subsequently, single-component nonlinear solitary waves were investigated using other nonlinear equations, such as the Korteweg-de Vries equation, the nonlinear Schr\"odinger equation, the Benjamin-Bona-Mahony equation, the sine-Gordon equation, and many others.

The two-component nonlinear solitary wave is a bound state of two single-component nonlinear solitary waves with the same velocities and parallel or mutual perpendicular polarisations, for instance, for waveguide modes. In general case, each of these components has different frequencies and wave numbers. Such pulses are sometimes called breather molecules [11-16]. The two-component vector breathers are of particular interest because they are used in many different physical phenomena.

The two-component vector breather is the bound state of the single-component breather pair with parallel polarisation. The first breather oscillates with the sum, and the second with the difference of the frequencies and wave numbers (SDFW). Such vector breather is characterised by propagating a two-component nonlinear optical or acoustic pulse in resonantly absorbing media. It was first investigated using the generalised perturbative reduction method under the effect of self-induced transparency and has been called the vector $0\pi$ pulse [14-16]. Subsequently, such a pulse was considered in many well-known nonlinear differential equations, such as the sine-Gordon equation, the nonlinear Schr\"odinger equation, the Maxwell equation in dispersive and Kerr medium, the different versions of the Benjamin-Bona-Mahony equations, the Hirota equation, the Born-Infeld equation, the Born-Infeld-sine-Gordon equation, the Boussinesq equation, the Boussinesq-type equation, the Maxwell-Bloch system of equations, the system of the elastic wave equation and magnetic Bloch equations, the generalized nonlinear wave equation, etc. These equations describe completely different nonlinear phenomena in absolutely different physical systems for various fields of physics, such as condensed matter physics, fluid dynamics, plasma, field theory, and so on (see, for instance, [17-23] and references therein). This indicates that the two-component vector $0\pi$ pulse expresses universal properties of matter independent from their concrete physical nature.

All nonlinear partial differential equations investigated so far, in which two-component vector $0\pi$ pulses were considered, contain partial derivatives concerning spatial coordinates and time, including the first to the fourth order. However, there are equations describing important physical phenomena containing partial derivatives with respect to spatial coordinates and time of a higher order. These include, in particular, nonlinear partial differential equations containing partial derivatives concerning spatial coordinates and time of the sixth order. It is noteworthy to express the sixth-order generalised Boussinesq-type equations among them. These equations are used to study the propagation of the nonlinear solitary water waves with surface tension [24-29].

The two versions of the sixth-order generalised Boussinesq-type equations are considered.

The first version of the sixth-order generalised Boussinesq-type nonlinear wave equation has the following form:
\begin{equation}\label{1}\nonumber
\alpha \partial_{t,t} E + \beta \partial_{z,z} E + \delta \partial_{z,z,z,z} E + \nu \partial_{z,z,t,t} E
+ S \partial_{z,z,z,z,t,t} E= F[E]_{z,z}
\end{equation}
or in another form
\begin{equation}\label{1e}
\alpha \frac{\pa^{2} E }{\partial t^2}  + \beta \frac{\pa^{2} E }{\partial z^2} + \delta \frac{\pa^{4} E }{\partial z^4}  + \nu  \frac{\pa^{4} E}{{\pa z^2}{\partial t^2}}
+ S  \frac{\pa^{6} E}{{\pa z^4}{\partial t^2}}= \frac{\pa^{2} F(E) }{\partial z^2},
\end{equation}
where  $E(z,t)$  is a real function of space coordinate $z$ and time $t$ and represents the nonlinear wave profile, while $\alpha$, $\beta$, $\delta$, $\nu$, and $S$ are arbitrary constants. $F(E) $ is an arbitrary nonlinear function.

The second version of the sixth-order generalised Boussinesq-type nonlinear wave equation has the form:
\begin{equation}\label{2e}
\tilde{\alpha} \frac{\pa^{2} E }{\partial t^2}  + \tilde{\beta} \frac{\pa^{2} E }{\partial z^2} +\tilde{ \delta }\frac{\pa^{4} E }{\partial z^4}  + \tilde{\nu}  \frac{\pa^{4} E}{{\pa z^2}{\partial t^2}}
+ \tilde{S}  \frac{\pa^{6} E}{\pa z^6}= \frac{\pa^{2} F(E) }{\partial z^2}.
\end{equation}
 where $\tilde{\alpha}$, $\tilde{\beta}$, $\tilde{\delta}$, $\tilde{\nu}$, and $\tilde{S}$ are arbitrary constants.

Eqs.(1) and (2) differ from each other in terms of the sixth-order partial derivatives concerning spatial coordinates and time. In equation (1), the sixth-order derivative term has the form $\frac{\pa^{6} E}{{\pa z^4}{\partial t^2}}$, while in equation (2), it has a different form  $\frac{\pa^{6} E}{\pa z^6}$.

Here and below, for simplicity, we use the same function $E(z,t)$ for both Eqs.(1)
and (2), where it will not look confusing.

The nonlinear term $F(E)$ in the sixth-order generalised Boussinesq-type equations (1) and (2) very often has a nonlinearity of the second-order or third-order form [25-29].
We consider situations when the nonlinear arbitrary function has the form:
\begin{equation}\label{fg}
F(E)=-G E^{3}.
\end{equation}
where $G$ is arbitrary constant.

For the analysis of Eqs.(1) and (2), the pulse width $T$ plays an important role. We can consider two types of pulses. The first are ultrashort pulses [30], and the second are relatively wider pulses, for which the inequality $\omega T >> 1$ is valid. Slowly varying envelope approximation is usually used for the last type of pulse [31-35],where $\omega$ is the carrier wave frequency. For such kind pulses, the function $E(z, t)$ can be represented as:
\begin{equation}\label{eq4}
E(z,t)=\sum_{l=\pm1}\hat{E}_{l}(z,t) Z_l,\;\;\;\;\;\;\;\;\;\;Z_l=e^{{il(k z -\om t)}}
\end{equation}
where $Z_{l}= e^{il(kz -\om t)}$ is the fast oscillating function, $\hat{E}_{l}$  is the slowly varying complex envelope function, which satisfies inequalities
\begin{equation}\label{swa}
 \left|\frac{\partial \hat{E}_{l}}{\partial t}\right|\ll\omega
|\hat{E}_{l}|,\;\;\;\left|\frac{\partial \hat{E}_{l}}{\partial z
}\right|\ll k|\hat{E}_{l}|,
\end{equation}
$k$ is the wave number of the carrier wave. For the reality of $E$, we assume that: $ \hat{E}_{+1}= \hat{E}^{*}_{-1}$.

The purpose of the present work is to consider the two-component vector breather solutions of the cubic sixth-order generalised Boussinesq-type equations (1) and (2), under the condition Eq.(3), using the generalised perturbative reduction method Eq. (8) when the function $E(z,t)$ satisfies the slowly varying envelope approximation, Eqs.(4) and (5).

The rest of this paper is organised as follows: Section II is devoted to the first version of the cubic sixth-order generalised Boussinesq-type equation (1) for slowly varying complex envelope function  $\hat{E}_{l}$ and using the generalised perturbation reduction method Eq.(8). We will transform Eq.(7) for the function  $\hat{E}_{l}$ to the coupled nonlinear Schr\"odinger equations (16) for auxiliary functions. Section III will present the explicit analytical expressions for the shape and parameters of the two-component nonlinear pulse. Section IV is devoted to the second version of the cubic sixth-order generalised Boussinesq-type equation (2) and similar to Section II, we will transform this equation to the coupled nonlinear Schr\"odinger equations for auxiliary functions. Finally, in Section V, we will compare solutions of the Eqs.(1) and (2), and discuss the obtained results.

\section{The generalised perturbative reduction method for the  nonlinear equation $\eqref{1}$}

Initially, we consider the first version of the cubic sixth-order generalised Boussinesq-type equation \eqref{1e}.

Substituting Eq.(4) into Eq.(1) we obtain connection between carrier wave frequency $\omega$ and wave number $k$ in the following form
\begin{equation}\label{dis1}
k^2 {\omega}^2 \nu -k^4 {\omega}^2 S-{\omega}^2 \alpha -k^2 \beta +k^4 \delta  =0,
  \end{equation}
and the nonlinear equation for envelope function $\hat{E}_{l}$:
\begin{equation}\label{s1}
\sum _{l} Z_{l}(
i l A_{1}  \frac{\pa \hat{E}_{l}}{\partial t}
+A_{2}  \frac{\pa^{2} \hat{E}_{l}}{\partial t^2}
+i l A_{3}  \frac{\pa \hat{E}_{l}}{\pa z}
+A_{4} \frac{\pa^{2} \hat{E}_{l}}{{\pa z}{\partial t}}
+i l A_{5}  \frac{\pa^{3} \hat{E}_{l}}{{\pa z}{\partial t^2}}
+ A_{6}  \frac{\pa^{2} \hat{E}_{l}}{\pa z^2}
+i l A_{7} \frac{\pa^{3} \hat{E}_{l}}{{\pa z^2}{\partial t}}
+A_{8} \frac{\pa^{4} \hat{E}_{l}}{{\pa z^2}{\partial t^2}}
$$$$
+i l A_{9}   \frac{\pa^{3} \hat{E}_{l}}{\pa z^3}
+ A_{10} \frac{\pa^{4} \hat{E}_{l}}{{\pa z^3}{\partial t}}
+ i l A_{11}  \frac{\pa^{5} \hat{E}_{l}}{{\pa z^3}{\partial t^2}}
+A_{12} \frac{\pa^{4} \hat{E}_{l}}{\pa z^4}
+i l A_{13} \frac{\pa^{5} \hat{E}_{l}}{{\pa z^4}{\partial t}}
+ A_{14} \frac{\pa^{6} \hat{E}_{l}}{{\pa z^4}{\partial t^2}})=\frac{\pa^{2} F(E) }{\partial z^2},
\end{equation}
where
\begin{equation}\nonumber
A_1=-2\omega (k^4 S + \alpha - k^2 \nu),
$$$$
A_2=k^4 S + \alpha - k^2 \nu,
$$$$
A_3= 2 k (2 k^2  {\om}^2 S+\beta -2 k^2 \delta -{\om}^2 \nu ),
$$$$
A_4=-4 k {\om} (2 k^2 S-\nu ),
$$$$
A_5=-2 k (2 k^2 S-\nu ),
$$$$
A_6= 6 k^2 {\om}^2 S+  \beta -6 k^2 \delta -{\om}^2 \nu,
$$$$
A_7= 2{\om} (6 k^2 S-\nu ),
$$$$
A_8=-(6 k^2 S-\nu ),
$$$$
A_9=-4 k    ({\om}^2 S-\delta ),
$$$$
A_{10}= 8 k {\om} S,
$$$$
A_{11}= 4 k S,
$$$$
A_{12}=-({\om}^2 S-\delta ),
$$$$
A_{13}=-2 {\om} S,
$$$$
A_{14}=S.
\end{equation}

In the next stage, we use the generalised perturbative reduction method [14-17], which makes it possible to transform the nonlinear equation (7) for the slowly complex envelope function $\hat{E}_{l}$ to the coupled nonlinear Schr\"odinger equations for auxiliary functions $f_{l,n}^ {(\alpha)}$.

To following of the generalised perturbative reduction method  the complex function  $\hat{E}_{l}$ can be represented as [14-17]
\begin{equation}\label{gprm}
\hat{E}_{l}(z,t)=\sum_{\alpha=1}^{\infty}\sum_{n=-\infty}^{+\infty}\varepsilon^\alpha
Y_{l,n} f_{l,n}^ {(\alpha)}(\zeta_{l,n},\tau),
\end{equation}
where $\varepsilon$ is a small parameter,
$$
Y_{l,n}=e^{in(Q_{l,n}z-\Omega_{l,n}
t)},\;\;\;\zeta_{l,n}=\varepsilon Q_{l,n}(z-v_{l,n} t),
$$$$
\tau=\varepsilon^2 t,\;\;\;
v_{l,n}=\frac{\partial \Omega_{l,n}}{\partial Q_{l,n}}.
$$

Such a representation allows us to separate from $\hat{E}_{l}(z,t)$ in the more slowly changing functions $f_{l,n}^ {(\alpha)}(\zeta_{l,n},\tau)$.

It is assumed that the parameters $\Omega_{l,n}$ and $Q_{l,n}$ and the complex function $f_{l,n}^{(\alpha)}$ satisfy the inequalities for any values of $l$ and $n$:
\begin{equation}\label{rtyp}\nonumber\\
\omega\gg \Omega_{l,n},\;\;k\gg Q_{l,n},\;\;\;
\end{equation}
$$
\left|\frac{\partial
f_{l,n}^{(\alpha )}}{
\partial t}\right|\ll \Omega_{l,n} \left|f_{l,n}^{(\alpha)}\right|,\;\;\left|\frac{\partial
f_{l,n}^{(\alpha )}}{\partial z }\right|\ll Q_{l,n} \left|f_{l,n}^{(\alpha )}\right|.
$$

The parameters $\Omega_{l,n}$ and $Q_{l,n}$ are characterize more slowly oscillations in time and space coordinate in comparison with carrier wave frequency $\omega$ and wave number $k$.

Furthermore, for the sake of simplicity, we omit $l$  and $n$ indexes for the quantities $v_{l,n}$,
$\Omega_{l,n}$ and $Q_{l,n}$  in equations where this will not be messy.

In order to obtain a two-component vector breather solution of Eq.(1), we will follow the procedure developed for the generalised perturbative reduction method in several works [14-23].

By substituting Eq.(8) into Eq.(7) we obtain the following equation
\begin{equation}\label{eqz}
\sum_{l=\pm1}\sum_{\alpha=1}^{\infty}\sum_{n=\pm 1}\varepsilon^\alpha Z_{l} Y_{l,n}[W_{l,n}
+ \varepsilon i J_{l,n}\frac{\partial }{\partial \zeta_{l,n} } - \varepsilon^2 i l h_{l,n}  \frac{\partial }{\partial \tau}
-\varepsilon^{2} Q^{2}_{l,n} H_{l,n}\frac{\partial^{2} }{\partial \zeta_{l,n}^{2}}+O(\varepsilon^{3})]f_{l,n}^{(\alpha)}=\frac{\pa^{2} F(E) }{\partial z^2},
\end{equation}
where
\begin{equation}\label{wjh}
W_{l,n}=l n {\Omega} A_{1} - {\Omega}^2 A_{2} - l n Q A_{3} + {\Omega} Q A_{4} +l n {\Omega}^2 Q A_{5} - Q^{2} A_{6} -l n {\Omega} Q^{2} A_{7} + {\Omega}^{2} Q^{2} A_{8}
$$$$
+l n Q^{3} A_{9}-{\Omega} Q^{3} A_{10}
-l n {\Omega}^{2} Q^{3} A_{11} + Q^{4} A_{12}+l n {\Omega} Q^{4} A_{13}-{\Omega}^{2} Q^{4} A_{14},
$$$$
J_{l,n}=l Q   A_{3} - n \Omega Q  A_{4} - l \Omega^{2} Q  A_{5} + 2 n Q^{2}  A_{6} +2 l \Omega Q^{2}
A_{7} - 2 n \Omega^{2} Q^{2}  A_{8} - 3 l Q^{3}  A_{9} +3 n \Omega Q^{3}  A_{10}
$$$$
+3 l \Omega^{2} Q^{3}  A_{11} - 4 n Q^{4} A_{12} - 4 l \Omega Q^{4}  A_{13} + 4 n \Omega^{2} Q^{4}  A_{14}
- l Q  A_{1} v + 2 n \Omega Q  A_{2} v - n Q^{2} A_{4} v
$$$$
 -2 l \Omega Q^{2}  A_{5} v +l Q^{3}  A_{7} v - 2 n \Omega Q^{3}  A_{8} v + n Q^{4}  A_{10} v+ 2 l \Omega Q^{4}  A_{11} v - l Q^{5}  A_{13} v + 2 n \Omega Q^{5}  A_{14} v,
$$$$
h_{l,n}=- A_{1} + 2 l n \Omega A_{2} - l n Q A_{4} -2 \Omega Q A_{5} + Q^{2} A_{7} - 2 l n \Omega Q^{2} A_{8} + l
n Q^{3} A_{10} + 2 \Omega Q^{3} A_{11} - Q^{4} A_{13} + 2 l n \Omega Q^{4} A_{14},
$$$$
H_{l,n}=-A_{6} +\Omega^{2} A_{8} +3 l n Q A_{9} -3 \Omega Q A_{10} -3 l n \Omega^{2} Q A_{11} +6 Q^{2} A_{12} +6 l n \Omega Q^{2} A_{13} - 6 \Omega^{2} Q^{2} A_{14}
$$$$
+ A_4 v +2 l n \Omega A_{5} v + 4 \Omega Q A_{8} v - 3 Q^{2} A_{10} v - 6 l n \Omega Q^{2} A_{11} v +4 l n Q^{3} A_{13} v -8 \Omega Q^{3} A_{14} v
$$$$
- A_{2} v^{2} + l n Q A_{5} v^{2} + Q^{2} A_{8} v^{2}-l n Q^{3} A_{11} v^{2}- Q^{4} A_{14} v^{2}-l n A_{7} (\Omega +2 Q v ).
\end{equation}

The nonlinear term $\frac{\pa^{2} F(E) }{\partial z^2}$ of the equation (9) is of order to $\varepsilon^{3}$.

We equate the terms with the same order of $\varepsilon$ to zero. In the first order of $\varepsilon$, we follow the trend when $ f_{l,n}^{(1)}\neq0 $ the connection between the parameters $\Omega_{l,n}$ and $Q_{l,n}$  has the form
\begin{equation}\label{diss1}
l n {\Omega} A_{1} - {\Omega}^2 A_{2} - l n Q A_{3} + {\Omega} Q A_{4} +l n {\Omega}^2 Q A_{5} - Q^{2} A_{6} -l n {\Omega} Q^{2} A_{7} + {\Omega}^{2} Q^{2} A_{8}
$$$$
 +l n Q^{3} A_{9}-{\Omega} Q^{3} A_{10}
-l n {\Omega}^{2} Q^{3} A_{11} + Q^{4} A_{12}+l n {\Omega} Q^{4} A_{13}-{\Omega}^{2} Q^{4} A_{14}=0.
\end{equation}

From Eq.(11) we obtain
\begin{equation}\label{vg}
v_{l,n} = \frac{1}{M}[l n ( A_{3} - {\Omega}^{2} A_{5} +2  {\Omega} Q A_{7}-3  Q^{2} A_{9} +3  {\Omega}^{2} Q^{2} A_{11} -4  {\Omega} Q^{3} A_{13}) -{\Omega} A_{4} + 2 Q A_{6}
$$$$
-2 {\Omega}^{2} Q A_{8}  + 3 {\Omega} Q^{2} A_{10}-4 Q^{3} A_{12}+ 4 {\Omega}^{2} Q^{3} A_{14}],
\end{equation}
where
$$
M={l n A_{1} -2 {\Omega} A_{2}+Q A_{4} +2 l n {\Omega} Q A_{5} -l n Q^{2} A_{7} +2 {\Omega} Q^{2} A_{8} - Q^{3} A_{10}-2 l n {\Omega} Q^{3} A_{11}+l n Q^{4} A_{13}-2 {\Omega} Q^{4} A_{14}}.
$$

Using  Eqs.(9), (10) and (11), in  the second order of $\varepsilon$, we obtain the equation $J_{l,n}=0$ for any values of indexes $l$ and $n$.

In the third order of $\varepsilon$,  the nonlinear equation (9) is reduced to the form
\begin{equation}\label{qw}
\sum_{l=\pm1}\sum_{n=\pm 1}\varepsilon^{3} Z_{l} Y_{l,n}[ - i l h_{l,n}  \frac{\partial }{\partial \tau}- Q_{l,n}^{2} H_{l,n}\frac{\partial^{2} }{\partial \zeta_{l,n}^{2}}]f_{l,n}^{(1)}=\frac{\pa^{2} F(E) }{\partial z^2}.
\end{equation}

Substituting Eqs.(8) and (4) into Eq.(3) for the nonlinear term proportional to $\varepsilon^{3} Z_{+1}$, we have:
\begin{equation}\label{non}
- 3  G[(k+Q_{+})^{2} ( | f_{+1,+1}^ {(1)}|^{2} +2 | f_{+1,-1}^ {(1)}|^{2} ) Y_{+1,+1} f_{+1,+1}^ {(1)}
$$$$
  +(k-Q_{-})^{2}   ( | f_{+1,-1}^ {(1)}|^{2}   +2   | f_{+1,+1}^ {(1)}|^{2} )Y_{+1,-1} f_{+1,-1}^ {(1)}].
\end{equation}
where
$$
 Q_{+}=Q_{+1,+1}= Q_{-1,-1},\;\;\;\;\;\;\;\;\;\;\;\;\;\;\;\;\;\;\;\;\;\;\;\;\;  Q_{-}=Q_{+1,-1}= Q_{-1,+1},
$$

Similarly, we can write a term proportional to $Z_{-1}$.

From Eqs. (13) and (14), in the third order of $\varepsilon$, we obtain the system of nonlinear equations
\textbf{}\begin{equation}\label{2eq}
  i \frac{\partial f_{+1,+1}^{(1)}}{\partial \tau} + Q_{+}^{2} \frac{H_{+1,+1} }{{h}_{+1,+1}} \frac{\partial^2 f_{+1,+1}^{(1)}}{\partial \zeta_{+1,+1} ^2}+\frac{3 G (k+Q_{+})^{2}}{ {h}_{+1,+1}}  ( | f_{+1,+1}^ {(1)}|^{2} + 2 | f_{+1,-1}^ {(1)}|^{2} ) f_{+1,+1}^ {(1)}=0,
$$$$
   i \frac{\partial f_{+1,-1 }^{(1)}}{\partial \tau} + Q_{-}^{2} \frac{H_{+1,-1} }{{h}_{+1,-1}} \frac{\partial^2 f_{+1,-1 }^{(1)}}{\partial \zeta_{+1,-1}^2}+\frac{3 G (k-Q_{-})^{2}}{{h}_{+1,-1}} ( |f_{+1,-1}^ {(1)}|^{2} +2 |f_{+1,+1} ^ {(1)}|^{2} )  f_{+1,-1}^ {(1)}=0.
 \end{equation}

\vskip+0.5cm
\section{The two-component vector breather }

After transformation to the variables $z$ and $t$, from Eqs.(15) we obtain the coupled nonlinear Schr\"odinger equations for the auxiliary functions $\Lambda_{\pm}=\varepsilon  f_{+1,\pm1}^{(1)}$ in the following form
\begin{equation}\label{pp2}
i (\frac{\partial \Lambda_{\pm}}{\partial t}+v_{\pm} \frac{\partial  \Lambda_{\pm}} {\partial z}) + p_{\pm} \frac{\partial^{2} \Lambda_{\pm} }{\partial z^{2}}
+q_{\pm}(|\Lambda_{\pm}|^{2}+2 |\Lambda_{\mp}|^{2} )\Lambda_{\pm}=0,
\end{equation}
where
\begin{equation}\label{pp4}
p_{\pm}=\frac{H_{+1,\pm 1} }{{h}_{+1,\pm 1}},
$$
$$
 q_{\pm}=\frac{3 G  (k\pm Q_{\pm })^{2}}{ {h}_{+1,\pm 1}},
$$
$$
v_{\pm }= v_{{+1,\pm 1}}.
\end{equation}

Various mathematical methods comprehensively study Eq.(16) and its solutions (see, for instance [36-39] and references therein).

Here, we will search for the solution of Eq.(16) in the form [14-17]
\begin{equation}\label{ue1}
\Lambda_{\pm }=\frac{R_{\pm }}{b T}Sech(\frac{t-\frac{z}{V_{0}}}{T}) e^{i(k_{\pm } z - \omega_{\pm } t )},
\end{equation}
where $R_{\pm },\; k_{\pm }$ and $\omega_{\pm }$ are the real constants, $V_{0}$ is the velocity of the nonlinear wave. We assume that
$k_{\pm }=\frac{V_{0}-v_{\pm}}{2p_{\pm}}<<Q_{\pm }$  and $\omega_{\pm }<<\Omega_{\pm },$
where
$
\Omega_{+}=\Omega_{+1,+1}= \Omega_{-1,-1},\;\;\;\; \Omega_{-}=\Omega_{+1,-1}= \Omega_{-1,+1}.
$

By combining Eqs. (4), (8) and (18), we obtain the two-component vector breather solution of the nonlinear wave equation (1) in the following form:
\begin{equation}\label{vb}
E(z,t)=\frac{2 }{b T}\sech(\frac{t-\frac{z}{V}}{T})\{ R_{+1} \cos[(k+Q_{+}+k_{+1})z -(\om +\Omega_{+}+\omega_{+1}) t]
+R_{-1}\cos[(k-Q_{-}+k_{-1})z -(\om -\Omega_{-}+\omega_{-1})t]\},
\end{equation}
where the connection between the width $T$ and the velocity $V_{0}$  of the two-component nonlinear pulse is determined as:
\begin{equation}\label{rrw}
T^{-2}=V_{0}^{2}\frac{v_{+}k_{+1}+k_{+1}^{2}p_{+}-\omega_{+1}}{p_{+}},
$$$$
{b}^{2}=V_{0}^{2} \frac{R_{+1}^{2}+2 R_{-1}^{2}}{2p_{+}}q_{+}.
\end{equation}

The connections between parameters of the nonlinear wave has the form:
\begin{equation}\label{ttw}
R_{+1}^{2}=\frac{p_{+}{q}_{-}-2 p_{-}q_{+}}{p_{-}{q}_{+}-2 p_{+}q_{-}}R_{-1}^{2},
\;\;\;\;\;\;\;\;\;
\omega_{+1}=\frac{p_{+}}{p_{-}}\omega_{-1}+\frac{V^{2}_{0}(p_{-}^{2}-p_{+}^{2})+v_{-}^{2}p_{+}^{2}-v_{+}^{2}p_{-}^{2}
}{4p_{+}p_{-}^{2}}.
\end{equation}

\vskip+0.5cm
\section{The generalised perturbative reduction method for the  nonlinear equation $\eqref{2e}$}

This section considers the second version of the cubic sixth-order generalised Boussinesq-type equation (2).
We will use the solution of Eq.(2) as we consider Eq.(1) in previous sections II and III.

In the first step, we substituted Eq.(4) into Eq.(2), and we obtained a connection between carrier wave frequency $\omega$ and wave number $k$ in the following form
\begin{equation}\label{dis12}
-k^6 \tilde{S}-{\omega}^2 \tilde{\alpha} -k^2 \tilde{\beta} +k^4 \tilde{\delta} +k^2 {\omega}^2 \tilde{\nu} =0
  \end{equation}
and the nonlinear equation for envelope function $\hat{E}_{l}$:
\begin{equation}\label{s2}
\sum _{l} Z_{l}(
i l B_{1}  \frac{\pa \hat{E}_{l}}{\partial t}
+B_{2}  \frac{\pa^{2} \hat{E}_{l}}{\partial t^2}
+i l B_{3}  \frac{\pa \hat{E}_{l}}{\pa z}
+B_{4} \frac{\pa^{2} \hat{E}_{l}}{{\pa z}{\partial t}}
+i l B_{5}  \frac{\pa^{3} \hat{E}_{l}}{{\pa z}{\partial t^2}}
+ B_{6}  \frac{\pa^{2} \hat{E}_{l}}{\pa z^2}
+i l B_{7} \frac{\pa^{3} \hat{E}_{l}}{{\pa z^2}{\partial t}}
$$$$
+B_{8} \frac{\pa^{4} \hat{E}_{l}}{{\pa z^2}{\partial t^2}}
+i l B_{9}   \frac{\pa^{3} \hat{E}_{l}}{\pa z^3}
+B_{10} \frac{\pa^{4} \hat{E}_{l}}{\pa z^4}
+i l B_{11} \frac{\pa^{5} \hat{E}_{l}}{\pa z^5}
+B_{12} \frac{\pa^{6} \hat{E}_{l}}{\pa z^6})=\frac{\pa^{2} F(E) }{\partial z^2},
\end{equation}
where
\begin{equation}\nonumber
B_1=-2 \omega (\tilde{\alpha} -k^2 \tilde{\nu} ),
$$$$
B_2=\tilde{\alpha} -k^2 \tilde{\nu},
$$$$
B_3= 2 k (3 k^4 \tilde{S} +\tilde{\beta} -2 k^2 \tilde{\delta} -{\omega}^2 \tilde{\nu} ),
$$$$
B_4= 4 k \omega \tilde{\nu},
$$$$
B_5=2 k \tilde{\nu },
$$$$
B_6= 15 k^4 \tilde{S}+\tilde{\beta} -6 k^2 \tilde{\delta} -{\omega}^2 \tilde{\nu},
$$$$
B_7=-2 \omega \tilde{\nu},
$$$$
B_8=\tilde{\nu},
$$$$
B_9= -4 k (5 k^2 \tilde{S}-\tilde{\delta}),
$$$$
B_{10}= -(15 k^2 \tilde{S}-\tilde{\delta}),
$$$$
B_{11}= 6 k \tilde{S},
$$$$
B_{12}=S.
\end{equation}

Using the generalised perturbative reduction method, we substitute the equation (8) into the equation (23) and obtain the following equation
\begin{equation}\label{s3}
\sum_{l=\pm1}\sum_{\alpha=1}^{\infty}\sum_{n=\pm 1}\varepsilon^\alpha Z_{l} Y_{l,n}[\tilde{W}_{l,n}
+ \varepsilon i \tilde{J}_{l,n}\frac{\partial }{\partial \zeta_{l,n} } - \varepsilon^2 i l \tilde{h}_{l,n}  \frac{\partial }{\partial \tau}
-\varepsilon^{2} Q^{2}_{l,n} \tilde{H}_{l,n}\frac{\partial^{2} }{\partial \zeta_{l,n}^{2}}+O(\varepsilon^{3})]f_{l,n}^{(\alpha)}=\frac{\pa^{2} F(E) }{\partial z^2},
\end{equation}
where
\begin{equation}\label{jwh2}
\tilde{W}_{l,n}=l n \Omega B_1 -  {\Omega}^2 B_2 -l n Q B_3 +  \Omega Q B_4 +l n {\Omega}^2 Q B_5 - Q^2 B_6
$$$$
 -l n \Omega Q^2 B_7+{\Omega}^2 Q^2 B_8 +l n Q^3 B_9 + Q^4 B_{10} -l n Q^5 B_{11} - Q^6   B_{12},
$$
$$
\tilde{h}_{l,n}=- B_1 + 2 l n \Omega B_2 - l n Q B_4 - 2  \Omega Q B_5 + Q^2 B_7 - 2 n l \Omega Q^2 B_8 ,
$$
$$
\tilde{J}_{l,n}=i Q \epsilon  (l B_3 - n \Omega B_4 - l \Omega^2 B_5 + 2 n Q B_6 + 2 l \Omega Q B_7 - 2 n \Omega^2 Q B_8 - 3 l Q^2 B_9 - 4 n Q^3 B_{10}
$$$$
 + 5 l Q^4 B_{11} + 6 n Q^5 B_{12}
- l B_1 v +2 n \Omega B_2 v -n Q B_4 v -2 l \Omega Q B_5 v +l Q^2 B_7 v -2 n \Omega Q^2 B_8 v),
$$
$$
\tilde{H}_{l,n}=B_6 +l n \Omega B_7 - \Omega^2 B_8 - 3 l n Q B_9 - 6 Q^2 B_{10} + 10 l n Q^3 B_{11} +15 Q^4 B_{12} $$$$
-B_4 v -2 l n \Omega B_5 v + 2 l n Q B_7 v - 4 \Omega Q B_8 v + B_2 v^2 - l n Q B_5 v^2 - Q^2 B_8 v^2.
\end{equation}

From  Eqs.(24) and (25) follows the connection between the parameters $\Omega_{l,n}$ and $Q_{l,n}$ for the cubic sixth-order  generalised Boussinesq-type equation (2) has the following form:
\begin{equation}\label{diss2}
l n \Omega B_1 -  {\Omega}^2 B_2 -l n Q B_3 +  \Omega Q B_4 +l n {\Omega}^2 Q B_5 - Q^2 B_6
$$$$
 -l n \Omega Q^2 B_7+{\Omega}^2 Q^2 B_8 +l n Q^3 B_9 + Q^4 B_{10} -l n Q^5 B_{11} - Q^6   B_{12}=0.
\end{equation}
From Eq.(26) we have the expression

\begin{equation}\label{vs2}
\tilde{v}_{l,n}=\frac{l n B_3 -\Omega B_4 -l n \Omega^2 B_5 + 2 Q B_6 + 2 l n \Omega Q B_7 - 2 \Omega^2 Q B_8 -3 l n Q^2 B_9 - 4 Q^3 B_{10} +5
l n Q^4 B_{11} + 6 Q^5 B_{12}}{l n B_1 - 2 \Omega B_2 + Q B_4 + 2 l n \Omega Q B_5 - l n Q^2 B_7 + 2 \Omega Q^2 B_8}.
\end{equation}

Similar to the previous sections for equation (1),  we can reduce the cubic sixth-order  generalised Boussinesq-type equation (2) to the coupled nonlinear Schr\"odinger equations and after solving obtain a two-component vector breather oscillating with SDFW Eq.(19). But the pulse parameters for Eq.(2) in Eq.(19) will be different compared to the solution for equation (1). In particular, it is necessary to make the following substitutions for the parameters of the nonlinear wave in the following equations (19), (20) and (21):
\begin{equation}\label{r}
p_{\pm} \rightarrow \frac{\tilde{H}_{+1,\pm 1} }{{\tilde{h}}_{+1,\pm 1}},\;\;\;\;\;\;
q_{\pm}\rightarrow \frac{3 G  (k\pm Q_{\pm })^{2}}{ {\tilde{h}}_{+1,\pm 1}},\;\;\;\;\;
v_{\pm }\rightarrow \tilde{v}_{{+1,\pm 1}}.
\end{equation}
where $v_{\pm }$ is determined from the Eq.(27).

Consequently, all equations obtained in Section III are also valid for Eq. (2) if we make the replacements concerning Eq.\eqref{r}.

\vskip+0.5cm

\section{Conclusion}

In this paper, we studied the two-component vector breather solutions for the two versions  of the cubic sixth-order generalised Boussinesq-type equations (1) and  (2). We considered nonlinear pulses with the width $T>>\Omega_{\pm }^{-1}>>\omega^{-1}$. Under the slowly varying envelope approximation Eqs.(4) and (5), we obtain the nonlinear wave equations (7) and (23)  for envelope function $\hat{E}_{l}$.

Using the generalised perturbation reduction method Eq.(8), the nonlinear wave equations (7) and (23)  for envelope function $\hat{E}_{l}$ were  transformed to the coupled nonlinear Schr\"odinger equations (16) for the auxiliary functions $\Lambda_{\pm 1}$.  As a result,
the two-component nonlinear pulse profiles for both versions of the cubic sixth-order  generalised Boussinesq-type equation (1) and  (2) we obtain  in the form of a vector breather Eq.(19) oscillating with the SDFW. The first  breather oscillates with the sum of frequencies $\omega-\Omega_{+}$ and wave numbers $k + Q_{+}$, and the second  breather oscillates with the difference of frequencies $\omega-\Omega_{-}$ and wave numbers $k - Q_{-}$.

The dispersion relations for  Eqs. (1) and  (2) are determined from Eqs.(6) and (22), respectively.
Meanwhile the connections between parameters $\Omega_{\pm}$ and $Q_{\pm}$  the Eqs. (1) and  (2)  are determined from Eqs.(11) and (26), respectively.

The other parameters of the nonlinear pulse for  Eq. (1) are determined from Eqs. (12),(17), (20), and (21)
and for  Eq. (2) the same equations are valid and we must make replacements as indicated in  Eq.(28). Comparison parameters of the nonlinear wave solution (19) for  Eqs. (1) and  (2), reveals that their parameters are absolutely different.

Consider a two-component nonlinear wave (19) solution of Eqs. (1) and  (2) that coincides with the vector $0\pi$ pulse of the self-induced transparency [14-17], which  has been investigated later in the set of well known equations (see, Section I ).

The study of two-component vector breathers for Eqs. (1) and  (2) expands the range of phenomena in which these nonlinear pulses can be formed. This is one more important confirmation that two-component vector breather Eq.(19) expresses general fundamental properties of matter.

In this paper, we have used the generalised perturbation reduction method Eq.(8) for the first time to solve equations that containing derivatives of the spatial coordinate and time of the sixth order. Thus, we have demonstrated the possibility of using this method to solve a wider range of equations.

\vskip+0.5cm

\end{document}